\documentclass[preprint2]{aastex}
\usepackage{amsmath}

\newcommand{\angstrom}{\text{\normalfont\AA}}

\shorttitle{Binding energies of small carbon clusters}
\shortauthors{Mauney et al.}

\begin{document}

\title{Formation and properties of astrophysical carbonaceous dust. I:
  {\it ab-initio} calculations of the configuration and binding
  energies of small carbon clusters}

\author{Christopher Mauney\altaffilmark{1,2}, Marco Buongiorno
  Nardelli\altaffilmark{3}, and Davide
  Lazzati\altaffilmark{1,2}} \affil{Oregon State University}
\email{mauneyc@onid.oregonstate.edu}

\altaffiltext{1}{Department of Physics, Oregon State University, 301
  Weniger Hall, Corvallis, OR 97331}

\altaffiltext{2}{Department of Physics, NC State University, 2401
  Stinson Drive, Raleigh, NC 27695-8202}

\altaffiltext{3}{Department of Physics, University of North Texas,
  Denton, TX 76203}

\begin{abstract} The binding energies of $n<100$ carbon clusters are
  calculated using the \emph{ab-initio} density functional theory code
  {\it Quantum Espresso}. Carbon cluster geometries are determined
  using several levels of classical techniques and further refined
  using density functional theory. The resulting energies are used to
  compute the work of cluster formation and the nucleation rate in a
  saturated, hydrogen-poor carbon gas. Compared to classical
  calculations that adopt the capillary approximation, we find that
  nucleation of carbon clusters is enhanced at low temperatures and
  depressed at high temperatures.  This difference is ascribed to the
  different behavior of the critical cluster size. We find that the
  critical cluster size is at $n=27$ or $n=8$ for a broad range of
  temperatures and saturations, instead of being a smooth function of
  such parameters. The results of our calculations can be used to
  follow carbonaceous cluster/grain formation, stability, and growth
  in hydrogen poor environments, such as the inner layers of
  core-collapse supernova and supernova remnants.
\end{abstract}


\section{Introduction}
Dust plays an important role at many levels in the observability, the
dynamics, and the evolution of astrophysical phenomena. Many heavy
atoms produced in stellar explosions are found in dust
\citep{Weingartner01}. The formation of planetary and stellar systems
is heavily affected by the abundance of dust \citep{Whittet03}; in
particular dust is the the building block of rocky planets in
protoplanetary disks \citep{Blum08}. Furthermore, dust in the ISM
scatters ultraviolet and visible light, and emits brightly in the
infrared \citep{Draine03}.

However, dust formation in stellar and galactic environments is not
well understood. The observation of crystalline silicates
\citep{Spoon06,Speck08} and dust formation in the hostile environment
of colliding winds of WR stars \citep{Williams90,Varricatt04}
underline serious holes in our understanding of the physics of the
phase transition at the base of dust formation.  In addition,
theoretical predictions of dust formation in supernova explosions
over-predict the amount of dust observed at early times by several
orders of magnitudes (on timescales of one year after core-collapse)
\citep{Todini01,Fallest11,Gall14}.  The limitations of the classical
theory of nucleation for dust formation have been repeatedly pointed
out \citep{Donn85,JonesA01,Lazzati08,Cherchneff09,Cherchneff10}. Among the weaknesses of the theory is that of assuming
quasi-equilibrium conditions for a strongly out of equilibrium
phenomenon, the assumption of local thermal equilibrium, ignoring the
presence of other components in the gas phase, and the capillary
approximation. The capillary approximation is used to predict the
binding energy of very small molecular clusters by assuming: (i) that
they are spherical in shape, (ii) that a surface can be defined for
them, and (iii) that the surface energy does not depend on the cluster
size. Since the nucleation rate depends exponentially on the energy of
cluster formation, and since the critical size of refractory grains is
usually small due to the large temperatures at which nucleation
occurs, the capillary approximation can be devastating to the accuracy
of the results. Another limitation of dust formation calculations is
that they assume homogeneous nucleation, i.e. that each compound
nucleated independently of all the other atoms and molecules in the
gas phase. While this is inaccurate in some environments such as
the interstellar medium, where abundant hydrogen can incorporate into dust materials, we here consider only pure carbon clusters. Our results should
be therefore considered relevant only for carbon-rich environments.

In this article we bypass the difficulties of the capillary approximation by using the atomistic formulation of the energy of cluster formation to determine nucleation rates. This approach accounts for clusters to form in irregular shapes, and has been shown to better model nucleation rates in crystals \citep{Kashchiev08}. Using modern
molecular and quantum mechanical methods we find the ground state
configuration and cohesive energy (and thus binding energy) of small carbon clusters up to
$n=99$ atoms.  While results for some clusters are available in the
literature (e.g., \cite{Jones99,Kent99}) we offer here a comprehensive
study of all sizes up to $n=99$, as required for application in the
nucleation theory. We use these data to make predictions of critical
cluster sizes and nucleation rates using both the thermodynamic and
the kinetic nucleation theory.

This paper is organized as follows: in Section \ref{sec:methods} we discuss the
semi-classical and density functional theory (DFT) methods adopted in
the calculations and in Section \ref{sec:dft_results} we present the resulting configurations
and cohesive energies. In Section \ref{sec:nucleation} we briefly introduce the classical
and kinetic nucleation theories, discuss where our results are
relevant, and apply them to compute nucleation rates. A discussion of
the whole paper is presented in Section \ref{sec:conclusion}.

\section{Methods}
\label{sec:methods}
\subsection{Determination of lowest-energy candidate configurations} \label{sec:geo_search}

Before DFT calculations can be carried out on the ground-state
configuration of carbon clusters, the geometric configuration of the
molecular cluster must be found to a good approximation. The methods
below are designed to efficiently search the Potential Energy Surface
(PES) for the increasingly complex space of possible arrangements.
The computational requirements for a PES search using \emph{ab-initio} quantum methods are quite
demanding. We therefore use a simpler
empirical bond-order potential to search for ground state candidate
geometries. These candidates will be further optimized during the density functional
calculation before the final determination of ground-state energy.

\subsubsection{Brenner Potential}
\label{sec:brenner_potential}

The simplified bond-order hydrocarbon potential of Brenner
\citep{Brenner90} is used in candidate search. Binding potential
between any two atoms in the cluster is written as a sum between
attractive and repulsive forces:
\begin{equation}
  \label{eq:brenner} E_{ij} = V_R(r_{ij}) - \bar{B_{ij}}V_A(r_{ij})
\end{equation}
giving the total binding energy of the cluster as:
\begin{equation}
\label{eq:brenner_eb}
E_b = \frac{1}{2} \sum_{i,j} E_{ij} = \sum_i \sum_{j > i}
\left[V_R(r_{ij}) - \bar{B_{ij}}V_A(r_{ij})\right]
\end{equation}

The repulsive and attractive pair potentials are given by:
\begin{equation} \label{eq:brenner_vr}
V_R(r_{ij}) = f_{ij}(r_{ij})\frac{D^{(e)}}{S - 1} \exp^{-\sqrt{2S}\beta(r_{ij}-R^{(e)})}
\end{equation}
and
\begin{equation} \label{eq:brenner_va}
V_A(r_{ij}) = f_{ij}(r_{ij})\frac{D^{(e)}S}{S - 1} \exp^{-\sqrt{2/S}\beta(r_{ij}-R^{(e)})}
\end{equation}

Defining the smooth pairwise cutoff function \(f_{ij}(r_{ij})\) as
\begin{equation} \label{eq:brenner_fij}
f_{ij}(r) =
\begin{cases}
  1 & \text{if } r_{ij} \le R^{(1)} \\
  \frac{1}{2}\left[1 + \cos \frac{\pi(r - R^{(1)})}{(R^{(2)} - R^{(1)})}\right] & R^{(1)} \le r_{ij} \le R^{(2)}\\
  0 & r_{ij} \ge R^{(2)}
\end{cases}
\end{equation}

The bond-order term \(\bar{B}_{ij}\) is given as

\begin{equation} \label{eq:brenner_bbij}
  \bar{B}_{ij} = \frac{B_{ij} + B_{ji}}{2}
\end{equation}

\begin{equation} \label{eq:brenner_bij}
  B_{ij} = \left( 1 + \sum_{k \ne i,j} G(\theta_{ijk})f_{ij}(r_{ik})\right)^{-\delta}
\end{equation}

\begin{equation} \label{eq:brenner_gijk}
  G(\theta_{ijk}) = a_0 \left(1 + \frac{c_0^2}{d_0^2} - \frac{c_0^2}{d_0^2 + (1 + \cos \theta_{ijk})}\right)
\end{equation}

The only variables in Eqs.(\ref{eq:brenner} -
\ref{eq:brenner_gijk}) are the inter-atomic distance \(r_{ij}\) and
angle \(\theta_{ijk}\). All other terms are parameters, listed in
Table \ref{tbl:brenner_param_table}. Values are taken from
\citep{Brenner90}.

\begin{deluxetable}{rrrrrrrrrr}
\tablecolumns{8}
\tablewidth{0pc}
\tablecaption{Brenner potential parameters}
\tablehead{
\colhead{$D^{(e)}$} & \colhead{$S$}   & \colhead{$\beta$} & \colhead{$R^{(e)}$} & \colhead{$\delta$} & \colhead{$a_0$} &
\colhead{$c_0$} & \colhead{$d_0$} &
\colhead{$R^{(1)}$} & \colhead{$R^{(2)}$} }
\startdata
6.325 & 1.29 & 1.5 & 1.315 & 0.80469 & 0.0011304 & 19 & 2.5 & 1.7 & 2.0\\
\enddata
\label{tbl:brenner_param_table}
\end{deluxetable}

\subsubsection{Cluster Geometry Search}
The computation of a geometric configuration of molecular clusters in
the ground state is the subject of much ongoing research
(e.g.\cite{Johnston02,Bauschlicher10,Goumans12}).  In the following sections we briefly
describe the array of methods we utilized in locating minimum energy
structures.  All methods are statistical in nature, and no single
result of any of the algorithms below suffices as a guaranteed minimal
state.  As these are the configurations found from a semi-classical
empirical potential, the absolute minima as well as several
near-degenerate configurations are chosen as candidates for DFT
calculations.

\paragraph{Simulated Annealing}
Simulated annealing is a Monte-Carlo search algorithm that moves the
positions of atoms randomly in small increments.  A
\emph{cost-function} describes the favorability of the configuration.
In this work the cost-function is the Brenner potential
(Section \ref{sec:brenner_potential}).  The algorithm also applies a
"temperature" to the configuration to encourage the escape of local
minima.

Every move that lowers the cost-function is accepted. Moves that
increase the value of the cost-function are subject to the Metropolis
criteria
\begin{equation} \label{eq:sa_dist0}
e^{-\Delta E / T} > \alpha
\end{equation}
where $\alpha$ is a random number on the range $[0,1)$.  After some
number of moves fine-tuned for the particular search, the temperature
is lowered as
\begin{equation} \label{eq:sa_temp}
T' = cT
\end{equation}
where $c < 1.0$ is the factor that controls cooling, usually set
between $0.8 \leq c \leq 0.99$.  These steps are repeated until the
temperature falls below a threshold value, typically near zero,
indicating the termination of the algorithm.

\paragraph{Basin Hopping}
Basin hopping \citep{Wales97} is a another Monte Carlo algorithm
similar to simulated annealing.  Atomic positions are moved at random
and a cost-function used to for determining optimal configurations.
The algorithm is also given a temperature and uphill moves are again
accepted according to the Metropolis criteria (\ref{eq:sa_dist0}).

Basin hopping introduces into this procedure a cost-function
minimization at each generation of new configurations.  After each
move step, a local geometric optimization (conjugate-gradient,
Broyden-Fletch-Goldfarb-Shanno, ect) is applied on atomic positions.
\begin{equation}
\label{basin_hopping_e}
\hat{E} = \min{E(\mathbf{x})}
\end{equation}
The energy surface is thus transformed from a continuous landscape of
hills and valleys into a discrete stair-step, making "down-funnel"
trajectories in the search much quicker.  No "cooling" is applied to
the temperature as is done in simulated annealing.  Each generated
configuration is "frozen" by the minimization step, so the cooling of
the temperature is not necessary.

\paragraph{Minima Hopping}
In contrast to the previous methods, minima hopping
\citep{Goedecker04} does not generate new configurations based on
random moves, but rather smoothly follows the energy surface by
applying molecular dynamics to the system.  Starting from an initial
state, the system is given a kinetic energy and allowed to evolve
according to the equations of motion.  The stopping criterion of the
molecular dynamics algorithm is passing over a small (one or two)
number of hills on the energy surface (that is, the system goes over
from increasing to deceasing energy).

Like basin hopping, the new configurations are subjected to an
optimization after the molecular dynamics step.  After the
optimization step, the configuration is compared to the previous one.
If the two configurations are determined to be the same (in terms of
atomic positions), the algorithm returns to the molecular dynamics
step with the previous configuration.  Kinetic energy for the
molecular dynamics is increased to encourage the system to escape the
current energy well.  If the system escapes the current energy well
into a new unique one, the kinetic energy is reduced.

Minima hopping maintains an internal list of previously discovered
configurations, and moves that repeat these configurations are
rejected.  During the stage of attempting escape from the local energy
well, if an escape is made into a previously located minima the
kinetic energy is increased to encourage exploration of other sites.
Similar to the previous methods, minima hopping allows an energy
increase when moving between neighboring minima.  The algorithm
maintains an adaptive energy threshold, where new configurations with
an energy increase below this threshold are accepted.  Moves which
energy increase below this threshold are rejected.  All moves to
unique configurations that decrease the energy are accepted.  When a
move is accepted, this threshold is made smaller, while rejected move
lead to an increased threshold.

\subsubsection{Combined Approach}
The methods presented above provide a sufficient toolkit for searching
for minimal energy configurations of carbon clusters.  We use all
three in an overlapping and reinforcing manner, where the results of
one method are fed into another.  Simulated annealing and basin
hopping with broad moves are used (for $n < 20 $ and $n \geq 20$
respectively) to generate starting configurations for minima hopping,
whose results are then subjected to simulated annealing and basin
hopping with again with finer moves.  This process is repeated until
further refinement yields no new structures.

This work uses the Broyden-Fletcher-Goldfarb-Shanno method (BFGS) as
the minimizer for both basin hopping and minima hopping optimization
steps.  Parameters for initial minima hopping runs are those given in
\citep{Goedecker04}, but these are adjusted to refine the current
search should further runs of minima hopping be necessary.  Molecular
dynamics are done with Verlet integration with a time-step of $\Delta
t = 2$ fs.

\subsubsection{Seeding and Large odd-numbered clusters}
Small clusters ($n < 20$) are randomly seeded.  In this regime the
search space is small enough that we need only apply simulated
annealing to locate suitable configurations for DFT analysis.  As the
size of the search space increases, it becomes very advantageous for
each algorithm to provide a reasonable seed to initiate the search,
rather than random starting positions.

For clusters $n \leq 60$, search seeds are randomly distributed on a
spherical surface.  The radius of the surface is chosen to be large
enough that each random position is sufficiently (approximately half an
average C-C bond length) far away from all other positions.  For
clusters with $n > 60$ atoms searching for the most stable
configurations from random initial positions becomes computationally
expensive.  It is preferable to begin the search with an educated
guess, guided by physical considerations.  In particular, the
stability of 3-coordinated carbon clusters is well known, and the
emergence of fullerene and fullerene-like cages in the results $n \leq
60$ from random initial seeds indicates that starting from a fullerene
structure will accelerate the search process.  For even-numbered
clusters with $n > 60$, we generate the initial seed using the
fullerene generation algorithm given in \citep{Brinkmann97}.  The full
range of search methods are applied on these initial seeds.

Odd-numbered clusters with $n > 60$ cannot form a proper fullerene
configuration.  Starting from an even-numbered fullerene, the maximum
bonding that can take place with an extra atom is a 2-coordinated
dangling bond somewhere near the fullerene surface.  In searching for
odd-numbered fullerene configurations, we seed the search with the
previously found even-numbered fullerene along with an extra atom
placed at a random position half an average bond length above the
surface of the fullerene.  The system is is then minimized.  The
minimized configuration is stored, and the process is repeated at a
new location for the appended atom.  The configuration with the lowest
energy is taken as the optimal configuration. We stress that even if
the search algorithm is seeded with a fullerene-like configuration, a
full search for the optimal configuration is performed allowing for
the possibility of non fullerene-like structures to be proofed.

\subsection{DFT calculation of binding energies} \label{sec:dft_calc}

\subsubsection{Density Functional Theory}
The candidate configurations from the searches above are then used as the initial configurations in DFT calculations.  Here outline density functional theory.

The Hamiltonian of a system of \(N\)-interacting electrons and fixed
nuclei is given by
\begin{equation} \label{eq:dft_hamiltonian}
\begin{split}
  H &= T + V_{ext} + W\\
  &= - \frac{\hbar ^2}{2 m_e} \sum_{i=1}^N \nabla ^2 + \sum_{i=1}^N v_{ext} (\mathbf{r}_i) + \frac{e^2}{2} \sum_{i \neq j} \frac{1}{|\mathbf{r}_i - \mathbf{r}_j|}
\end{split}
\end{equation}
\(T\) is the kinetic energy operator, and \(v_{ext}\) is some
external potential, in this context provided by the nucleus. The
many-body eigenstate \(\Psi_k\) satisfies
\begin{equation} \label{eq:dft_schrodinger}
\begin{split}
H \Psi_k (\mathbf{x}_1, \mathbf{x}_2, \cdots, \mathbf{x}_N) &= E_k \Psi_k (\mathbf{x}_1, \mathbf{x}_2, \cdots, \mathbf{x}_N)\\
H | \Psi_k \rangle &= E_k | \Psi_k \rangle
\end{split}
\end{equation}
where we have used the space-spin coordinate \(\mathbf{x} = (\mathbf{r}, \sigma)\).

Eq.(\ref{eq:dft_schrodinger}) is difficult to compute using
direct methods. Instead, in DFT the problem is approached by elevating
the number density \(n(\mathbf{r})\) from an observable of the system
to the fundamental function. That is, DFT inverts the relation

\begin{equation} \label{eq:dft_nden}
n(\mathbf{r}) = \langle \Psi | \hat{n}(\mathbf{r}) | \Psi \rangle
\end{equation}

The Hohenberg-Kohn (HK) theorems \citep{HK64, MartinTB} allows the wave function, in particular the ground
state wave function, to be expressed as a functional of the number
density

\[
| \Psi_0 \rangle = | \Psi_0 [n] \rangle = | \Psi [n_0] \rangle
\]

All observables determined from the wave function can be
derived from the number density, and themselves become functionals. Of
special interest is the ground state energy
\begin{equation}
\begin{split}
\langle \Psi_0[n] | H | \Psi_0[n] \rangle &= \min_{\Psi \rightarrow n_0} \langle \Psi | H | \Psi \rangle\\
&= \min_{\Psi \rightarrow n_0} \langle \Psi | T + W | \Psi \rangle + \int d\,\mathbf{r} v_{ext}(\mathbf{r})n_0(\mathbf{r})\\
&= F[n_0] + \int d\,\mathbf{r} v_{ext}(\mathbf{r})n_0(\mathbf{r})
\end{split}
\end{equation}
where the kinetic and repulsive coulomb terms have been combined into
the \textit{universal form} \(F[n]\), so called because it takes the
same functional form for any number of electrons.

As a result of the HK theorems, the ground state energy functional
\(E[n]\) is stationary with respect to the number density.
Determination of the ground state \(n_0 (\mathbf{r})\) is then a solution to the
constrained variational equation

\begin{equation} \label{eq:dft_numberdens}
\frac{\delta}{\delta n(\mathbf{r})} \left[ E[n(\mathbf{r})] - \mu \left(\int d\, \mathbf{r} n(\mathbf{r}) - N \right) \right] = 0
\end{equation}

The Kohn-Sham (KS) \emph{anstaz} provides a starting point for solving the above equation \citep{KohnSham65}.  The exact ground state density of an interacting system of particles can also be the ground state density of a set of non-interacting particles.  That is, the problem of solving for the many-body wave function \(|\Psi_0 \rangle\) is replaced by the problem of solve for many independent orbitals \(|\psi_i \rangle\).

The advantages of this method are intuitively clear.  Although the orbitals themselves express independent particles, the number density \(n_0(\mathbf{r})\) retains the details of many-body interactions through the exchange-correlation functional \(E_{xc}[n]\). For details on the formulation of the KS system, see \citep{MartinTB}.

\subsubsection{Quantum Espresso}
Quantum Espresso (QE) is an open-source DFT code that uses plane-wave
basis expansion for the orbitals in the Kohn-Sahm system
\citep{Giannozzi09}. Candidate configurations from the methods
discussed previously are used as the input configurations in QE. The
system is then subjected to a local optimization using the BFGS
algorithm.

QE uses an iterative self-consistent technique for solving the coupled
KS equations. Self-consistency is achieved when the relative
convergence is less than \(10^{-6}\). Mixing of the iterative number
densities is done using Broyden mixing with a mixing coefficient of
\(\alpha = 0.7\).

Electrons near the core of the atom are tightly bound, and are nearly
inert in bonding environments. Furthermore, electronic wavefunctions
peak sharply near the core, requiring a large number of plane-waves
for accurate treatment. It is useful to replace the full problem of
the central atom + all electrons with a pseudoatom + valence
electrons. The pseudoatom exerts a weaker pseudopotential than the
full Coulomb potential near the core, but is constructed to
sufficiently approach the all-electron potential beyond a critical
distance.  An ultrasoft pseudopotential is chosen to lessen the number
of plane-waves necessary for convergence. We use the convention of
\(E_{rho} = 12 \times E_{cut}\). Convergence in energy is achieved
when \(\Delta E_{tot} < 10^{-3}\).

Supercell optimization is necessary to negate any spurious electronic
interaction arising from the periodic boundary conditions in the
plane-wave treatment of QE. By implanting sufficient vacuum between
the molecule and the cell boundaries - our tests show that a vacuum of
at least 6 $\angstrom$ is necessary - the strength of the periodic
interactions drop off sufficiently quickly to allow investigation of
isolated systems.

\subsubsection{Selection of energy functionals}
Many-body exchange and correlation effects in DFT are captured by the
exchange-correlation energy functional \(E_{xc}[n(r)]\). This
functional, for most systems of interest, is not known exactly and
must be approximated. In QE, this functional is integrated with the
pseudopotentials, allowing the selection of the pseudopotential to
define the particulars of the DFT calculation. Investigation into the
behavior of carbon clusters, especially in the critical
ring-to-fullerene transition region, lead to the selection the
generalized gradient approximation (GGA) of Perdew-Burke-Ernzerhof
(PBE) for the functional of our DFT calculations \citep{Perdew96}.

\section{Results} \label{sec:dft_results}
\subsection{Overview}
The binding energy of an atomic cluster is given as the energy
necessary to remove the atoms from the cluster to infinity.  With
$E_T$ as the total system energy calculated with QE, and $E_1$ as the
ground-state energy of the isolated carbon atom, the binding energy of
a cluster of $n$ carbon atoms is given by

\begin{equation} \label{eq:bnd_e}
E_b = n E_{1} - E_T
\end{equation}

Our goal is to determine $E_b$. The results in this section, however,
are given as the cohesive energy $E_c = E_b / n$ to better illustrate
the properties of various clusters and to better allow for
straightforward comparison of similar work (see
Table \ref{tbl:compare}).

\subsection{C$_2$ to C$_9$}
\begin{figure}
\epsscale{1.0}
\plotone{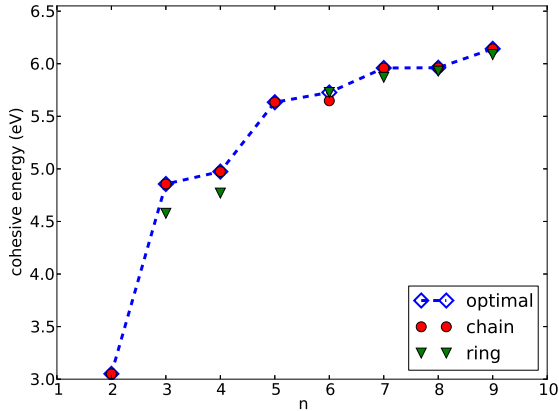}
\caption{Cohesive energies of carbon clusters C$_2$ to C$_9$}
\label{fig:c2c9}
\end{figure}

As shown in Fig. \ref{fig:c2c9}, all carbon structures in the range
C$_2$ to C$_9$, save for the strongly aromatic C$_6$, take the form of
linear chains.  The practicality of producing free medium-to-large
scale \( n > 5\) carbon chains, without terminal elements like
hydrogen for stabilization, is an open question \citep{Ravagnan02}.
Our calculations of C$_5$ indicates that the cyclic structure is
inherently unstable; our force relaxation procedure would always
unwind the cluster back into a linear chain, indicating that the
cyclic structure of C$_5$ is on or near to a saddle point on the
energy surface.  Optimal geometries and cohesive energies are given in
Table \ref{tbl:c2c9}

C$_6$ is the first molecule to demonstrate strong Huckle aromaticity,
the first molecule of the Huckle Rule pattern stating that the cyclic
structure for systems with a multiple of $4n + 2$ electrons are the most
energetically stable \citep{McEwen86}. There is general agreement in
theory and experiment that the distorted hexagon structure of C$_6$ is
the most stable \citep{Jones99}.

Although $n=8$, $n=9$ chain geometries are optimal, they are nearly
degenerate with the ring geometries.  In the range of temperatures
considered for nucleation (section \ref{sec:nucleation}), they are
both within one $kT$.  It is reasonable to conclude that the
transition from chains to rings can occur at $n=8$.

\begin{deluxetable}{rcc}
\tablecolumns{3}
\tablewidth{0pc}
\tablecaption{Optimal cohesive energies for C$_2$ to C$_9$}
\tablehead{
\colhead{$n$} & \colhead{Geometry} & \colhead{$E_c$ (eV)}}
\startdata
2 & chain & 3.051\\
3 & chain & 4.856\\
4 & chain & 4.974\\
5 & chain & 5.634\\
6 & ring & 5.727\\
7 & chain & 5.960\\
8 & chain & 5.962\\
9 & chain & 6.142\\
\enddata
\label{tbl:c2c9}
\end{deluxetable}

\subsection{C$_{10}$ to C$_{20}$}
\begin{figure}
\epsscale{1.0}
\plotone{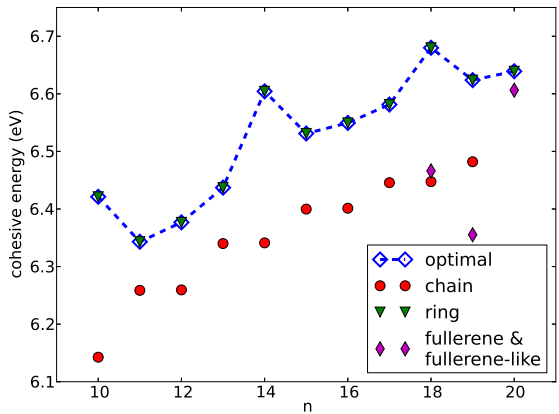}
\caption{Cohesive energies of carbon clusters C$_{10}$ to C$_{20}$}
\label{fig:c10c20}
\end{figure}

The strain due to the curvature of carbon rings for \(n \ge 10 \) is
sufficiently small that the extra C-C bond of the chain terminals
becomes energetically preferable. The transition between aromatic
states dominates in this regime, with the $4n + 2$ cyclic rings
showing pronounced peaks in stability (Fig.\ref{fig:c10c20}). The
\(D_{nh}\) symmetry of the classical calculation is broken into a
lower \(D_{n/2 h}\) due to Jahn-Teller distortion \citep{Saito99}.
The chain clusters show a repeating stair-step pattern, where
stability is unchanged between an even and the preceding odd-numbered
chain. The first fullerene-like states starting at C$_{18}$ are not as
energetically stable as the rings. Optimal geometries and cohesive
energies are given in Table \ref{tbl:c10c20}.

\begin{deluxetable}{rcc}
\tablecolumns{3}
\tablewidth{0pc}
\tablecaption{Optimal cohesive energies for C$_{10}$ to C$_{20}$}
\tablehead{
\colhead{$n$} & \colhead{Geometry} & \colhead{$E_c$ (eV)}}
\startdata
10 & ring & 6.421\\
11 & ring & 6.343\\
12 & ring & 6.377\\
13 & ring & 6.437\\
14 & ring & 6.604\\
15 & ring & 6.531\\
16 & ring & 6.549\\
17 & ring & 6.581\\
18 & ring & 6.680\\
19 & ring & 6.624\\
20 & ring & 6.639\\
\enddata
\label{tbl:c10c20}
\end{deluxetable}

\subsection{C$_{21}$ to C$_{30}$} \label{sec:dft_results_C21_C30}

The exact transition from cyclic carbon to fullerenes or
fullerene-like cages (hereafter fullerene-like) is an open
question. No experimental evidence is available that conclusively
shows where in this critical region the transition takes place, and
different levels of theory are in disagreement
\citep{Handschuh95,Helden93}.

\begin{figure}
\epsscale{1.0}
\plotone{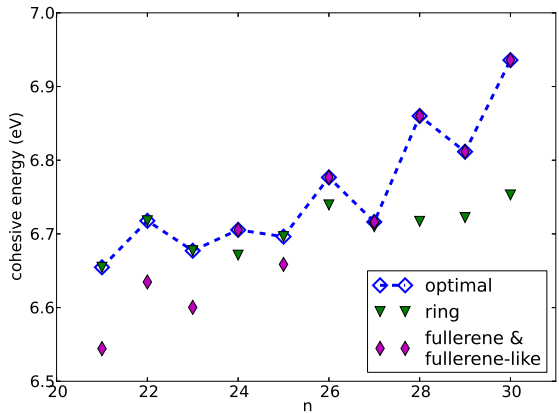}
\caption{Cohesive energies of carbon clusters C$_{21}$ to C$_{30}$}
\label{fig:c21c30}
\end{figure}

Fig. \ref{fig:c21c30} shows an initial continuation of the cyclic
rings through the final aromatic structure C$_{22}$. The fullerene
C$_{24}$ becomes the most stable configuration, although the
odd-numbered C$_{25}$ fullerene-like suffers in stability from a
dangling bond that returns the cyclic C$_{25}$ as the most stable
configuration.  C$_{27}$ is nearly degenerate in ring and
fullerene-like geometries.  The calculations performed here are in
good agreement on the transition point with other studies using
similar techniques \citep{Kent99,Jones99}. Compared to our
second-order approximation of exchange-correlation energy functional (PBE),
first-order approaches like local-density approximation (LDA) of the
exchange-correlation energy functional prefer an earlier transition to fullerenes,
whereas hybrid methods (B3LYP) prefer a later transition.

Continuing through particle number, the fullerenes and fullerene-like
quickly outpace the cyclic structures, a trend that continues for the
remainder of the clusters considered here. Optimal geometries and
cohesive energies are given in Table \ref{tbl:c21c30}.

\begin{deluxetable}{rcc}
\tablecolumns{3}
\tablewidth{0pc}
\tablecaption{Optimal cohesive energies for C$_{21}$ to C$_{30}$}
\tablehead{
\colhead{$n$} & \colhead{Geometry} & \colhead{$E_c$ (eV)}}
\startdata
21 & ring & 6.655\\
22 & ring & 6.718\\
23 & ring & 6.677\\
24 & fullerene & 6.705\\
25 & ring & 6.696\\
26 & fullerene & 6.777\\
27 & fullerene-like/ring & 6.716\\
28 & fullerene & 6.860\\
29 & fullerene-like & 6.812\\
30 & fullerene & 6.936\\
\enddata
\label{tbl:c21c30}
\end{deluxetable}

\subsection{C$_{31}$ to C$_{99}$} \label{sec:dft_results_C31_C99}
\begin{figure}
\epsscale{1.0}
\plotone{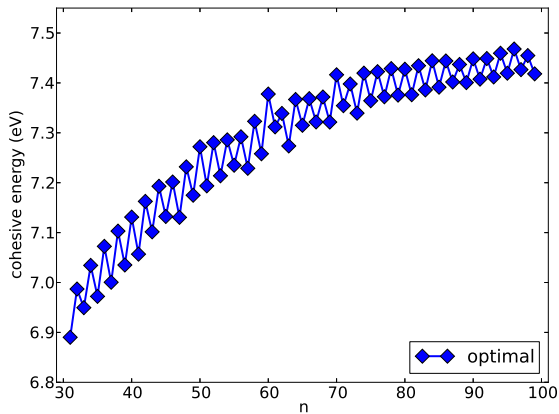}
\caption{Cohesive energies of carbon clusters C$_{31}$ to C$_{99}$}
\label{fig:c31c99}
\end{figure}

Fullerene and fullerene-like structures occupy the remainder of the
optimally stable carbon clusters (Fig.\ref{fig:c31c99}, Table \ref{tbl:c31c99}). The binding energy is monotonically increasing,
although there is a clear zig-zag pattern evident between the even-
and odd-numbered fullerene clusters.

\begin{figure}
\epsscale{1.0}
\plotone{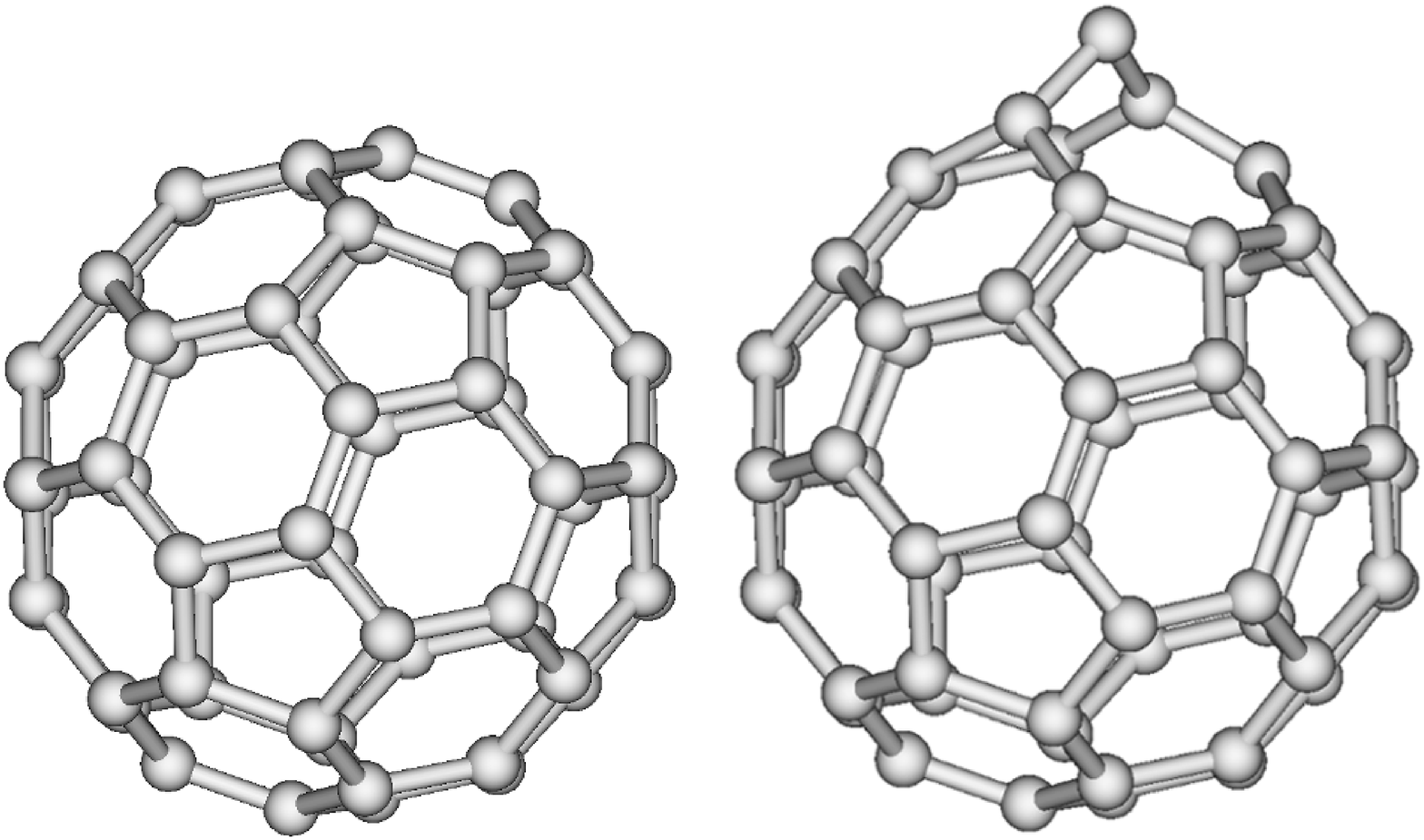}
\caption{\textit{left}: C$_{60}$ fullerene.  \textit{right}: C$_{61}$ fullerene-like
cage with appended atom on top.}
\label{fig:c60c61}
\end{figure}

Odd numbered fullerene-like clusters, shown in Fig.\ref{fig:c60c61},
are 2-coordinated at the appended atom, accounting for the clear
instability in relation to the even numbered clusters. This extra atom
prevents a fully closed cage structure associated with
fullerenes. These clusters have been observed in the experiments with
the fragmentation of C$_{60}$ \citep{Deng93,Kong01}, although they are
generally regarded as intermediate states.

Strong peaks in stability are present at both C$_{60}$ and C$_{70}$,
where the Isolated Pentagon Rule (IPR) is manifest \citep{Kroto87}.
The most significant strain in fullerenes is due to the sub-optimal
stability of the pentagon faces, which however are required to
complete the geometry of the fullerene. IPR fullerenes keep pentagon
faces maximally separated, minimizing the effect of pentagon strain
and yielding a larger stability relative to the surrounding clusters.

A comparison of our results with other theoretical investigations as
well as with experimental results is given in
Table \ref{tbl:compare}. The biggest differences are encountered when
comparing our results to theoretical DFT calculations adopting less
sophisticated pseudopotential (e.g., about 15\% difference for
C$_4$). Our results are comparable to within 1\% to other DFT results
adopting GGA pseudopotentials. Finally, comparison with experimental
results show that our results (and in general DFT results)
overestimate the cohesive energy by approximately 5\%. Overall the
agreement is quite good, even though the source of the systematic
difference with experimental results deserves further investigation.

\begin{deluxetable}{rccrccrcc}
\tablecolumns{9}
\tablewidth{0pc}
\tabletypesize{\scriptsize}
\tablecaption{Optimal cohesive energies for C$_{31}$ to C$_{99}$}
\tablehead{
\colhead{$n$} & \colhead{Geometry} & \colhead{$E_c$ (eV)} &
\colhead{$n$} & \colhead{Geometry} & \colhead{$E_c$ (eV)} &
\colhead{$n$} & \colhead{Geometry} & \colhead{$E_c$ (eV)}}
\startdata
31 & fullerene-like &  6.890 &  51 & fullerene-like &  7.194 &  71 & fullerene-like &  7.354\\
32 &  fullerene &  6.987 &  52 &  fullerene &  7.280 &  72 &  fullerene &  7.398\\
33 & fullerene-like &  6.950 &  53 & fullerene-like &  7.214 &  73 & fullerene-like &  7.339\\
34 &  fullerene &  7.034 &  54 &  fullerene &  7.286 &  74 &  fullerene &  7.420\\
35 & fullerene-like &  6.972 &  55 & fullerene-like &  7.235 &  75 & fullerene-like &  7.364\\
36 &  fullerene &  7.072 &  56 &  fullerene &  7.292 &  76 &  fullerene &  7.423\\
37 & fullerene-like &  7.000 &  57 & fullerene-like &  7.229 &  77 & fullerene-like &  7.372\\
38 &  fullerene &  7.103 &  58 &  fullerene &  7.323 &  78 &  fullerene &  7.429\\
39 & fullerene-like &  7.035 &  59 & fullerene-like &  7.258 &  79 & fullerene-like &  7.375\\
40 &  fullerene &  7.131 &  60 &  fullerene &  7.378 &  80 &  fullerene &  7.427\\
41 & fullerene-like &  7.057 &  61 & fullerene-like &  7.312 &  81 & fullerene-like &  7.376\\
42 &  fullerene &  7.163 &  62 &  fullerene &  7.339 &  82 &  fullerene &  7.434\\
43 & fullerene-like &  7.102 &  63 & fullerene-like &  7.274 &  83 & fullerene-like &  7.386\\
44 &  fullerene &  7.193 &  64 &  fullerene &  7.367 &  84 &  fullerene &  7.444\\
45 & fullerene-like &  7.132 &  65 & fullerene-like &  7.315 &  85 & fullerene-like &  7.391\\
46 &  fullerene &  7.201 &  66 &  fullerene &  7.369 &  86 &  fullerene &  7.444\\
47 & fullerene-like &  7.131 &  67 & fullerene-like &  7.322 &  87 & fullerene-like &  7.402\\
48 &  fullerene &  7.232 &  68 &  fullerene &  7.372 &  88 &  fullerene &  7.437\\
49 & fullerene-like &  7.175 &  69 & fullerene-like &  7.322 &  89 & fullerene-like &  7.401\\
50 &  fullerene &  7.272 &  70 &  fullerene &  7.416 &  90 &  fullerene &  7.448\\
51 & fullerene-like &  7.194 &  71 & fullerene-like &  7.354 &  91 & fullerene-like &  7.407\\
52 &  fullerene &  7.280 &  72 &  fullerene &  7.398 &  92 &  fullerene &  7.449\\
53 & fullerene-like &  7.214 &  73 & fullerene-like &  7.339 &  93 & fullerene-like &  7.412\\
54 &  fullerene &  7.286 &  74 &  fullerene &  7.420 &  94 &  fullerene &  7.459\\
55 & fullerene-like &  7.235 &  75 & fullerene-like &  7.364 &  95 & fullerene-like &  7.419\\
56 &  fullerene &  7.292 &  76 &  fullerene &  7.423 &  96 &  fullerene &  7.468\\
57 & fullerene-like &  7.229 &  77 & fullerene-like &  7.372 &  97 & fullerene-like &  7.427\\
58 &  fullerene &  7.323 &  78 &  fullerene &  7.429 &  98 &  fullerene &  7.455\\
59 & fullerene-like &  7.258 &  79 & fullerene-like &  7.375 &  99 & fullerene-like &  7.418\\

\enddata
\label{tbl:c31c99}
\end{deluxetable}

\begin{deluxetable}{rccr}
\tablecolumns{4}
\tablewidth{0pc}
\tablecaption{Comparison of Cohesive energies $E_c$ for selected clusters}
\tablehead{
\colhead{$n$} & \colhead{$E_c$ (this work)} & \colhead{$E_c$ (reference)} & \colhead{Method}}
\startdata
3 & 4.86 & 4.60\tablenotemark{a} & Exp\\
3 & 4.86 & 4.85\tablenotemark{f} & DFTB\\
4 & 4.97 & 4.75\tablenotemark{a} & Exp\\
4 & 4.97 & 5.75\tablenotemark{b} & DFT(LDA)\\
5 & 5.63 & 5.25\tablenotemark{a} & Exp\\
6  & 5.72 & 5.76\tablenotemark{b} & DFT(GGA)\\
24 & 6.71 & 6.46\tablenotemark{c} & DQMC\\
24 & 6.71 & 6.75\tablenotemark{b} & DFT(GGA)\\
28 & 6.86 & 6.63\tablenotemark{c} & DQMC\\
32 & 6.99  & 6.88\tablenotemark{d} & DFT(GGA)\\
60 & 7.38 & 7.40\tablenotemark{e} & DFT(LDA)\\
60 & 7.38 & 7.38\tablenotemark{g} & DFT(GGA)\\
60 & 7.38 & 7.07\tablenotemark{h} & Exp\\
70 & 7.42 & 7.42\tablenotemark{e} &DFT(LDA)\\
70 & 7.42 & 7.10\tablenotemark{g} & Exp
\enddata
\tablenotetext{a}{Ref. \cite{Drowart59}}
\tablenotetext{b}{Ref. \cite{Jones99}}
\tablenotetext{c}{Ref. \cite{Kent99}}
\tablenotetext{d}{Ref. \cite{Kobayashi92}}
\tablenotetext{e}{Ref. \cite{Saito91}}
\tablenotetext{f}{Ref. \cite{Menon93}}
\tablenotetext{g}{Ref. \cite{Perdew92}}
\tablenotetext{h}{Ref. \cite{Grimme97}}
\label{tbl:compare}
\end{deluxetable}

\section{Carbon dust nucleation} \label{sec:nucleation}

The principal application of the cohesive energies and structures
computed in this work is for the calculation of the nucleation rate of
carbonaceous dust in hydrogen-poor environments. The main example is
formation of graphitic or amorphous-carbon grains in the inner layers
of core-collapse supernovae, where the carbon to hydrogen number ratio
is typically in the range $10^6:1$ to $10^3:1$
\citep{Kozasa89,Kozasa91,Todini01,Fallest11}.

The deposition of carbon from the gas to the solid phase is a first
order phase transition.  The gas does not spontaneously go over from
the old to the new phase in bulk. Between the metastable and stable
phases there exists an energy barrier that makes a uniform and global
phase change (associated to a density change) highly improbable.  A
far more energetically favorable pathway is the random formation of
mirco-to-nanoscale density transition nuclei, a process known as
\textit{nucleation}.

The theory that follows utilizes the \textit{cluster model} of
nucleation. A bulk of $M$ particles exists in the gas phase, where a
cluster of $n \ll M$ particles at higher density forms within a small
region inside the bulk. In the following we consider both the
classical thermodynamic theory of nucleation and the kinetic theory of
nucleation.  For the classical theory we will use both the standard
capillary approximation (CNT-CAP) and our DFT-derived cohesive
energies in the atomistic formulation (ANT-DFT) for evaluating the work energy of cluster
formation. For the kinetic theory, we will use our DFT results for
deriving the cluster concentrations and the monomer detachment rates
(KNT-DFT).

\subsection{Classical nucleation theory}

In the classic nucleation theory, the rate at which new stable
clusters are nucleated in the supersaturated gas per unit volume and
time is given by:
\begin{equation} \label{eq:nucl_jrate0}
J = z \gamma S^* C_1^2 \sqrt{\frac{kT}{2 \pi m_0}}  e^{-W^* / kT}
\end{equation}
where $z$ is the Zeldovich factor (see below), $\gamma$ is the
sticking coefficient\footnote{The sticking coefficient $\gamma$ is the
  probability that an incoming monomer binds to the target cluster. It
  can in principle be different from unity and depend on the cluster
  size and on the gas temperature. However, we here follow the
  customary assumption of CNT and assume $\gamma=1$.} , $C_1$ is the
equilibrium gas density of carbon at the given temperature, $m_0$ is
the atomic mass of carbon, and $S^*$, $W^*$ are respectively the surface area and work of cluster formation, the star denoting that these quantities are associated with the cluster at the critical size (see below).

The work of cluster formation (WCF) is the energy necessary to form a
cluster in the new phase from the constituents of the old one. The WCF
of a cluster of size $n$ is given by
\begin{equation} \label{eq:nucl_work}
\begin{split}
W_n &= G_{new,n} - G_{old,n}\\
&=-n \Delta \mu + G_{ex,n}
\end{split}
\end{equation}
where $G$ is the Gibbs free energy and $\Delta \mu = k T \ln S$ is the
supersaturation \cite{Zettlemoyer69}.  T is the system temperature and
S denotes the supersaturation ratio of vapor-to-equilibrium pressure
$S = p/p_e$ (hereafter saturation).

Classical nucleation utilizes the capillary approximation to determine
the excess Gibbs free energy $G_{ex,n}$
\begin{equation} \label{eq:nucl_gex_cnt}
G_{ex,n} = \sigma_n S_n = \sigma_n c (n v_0)^{\frac{2}{3}}
\end{equation}
where $\sigma_n$ is the surface tension of the cluster,
$c=S_n/V_n^{2/3}$ is a shape factor\footnote{$c_s=(36\pi)^{1/3}$ for
  spherical grains.} and $v_0$ is the volume of occupied by a single
particle in the new phase. Clearly the capillary approximation is
inadequate in the regime where $n$ is small, and where a spherical
approximation for the molecular geometry is poor. Unfortunately,
it is at this scale that nucleation is most significant for
astrophysical refractory grains.

The capillary approximation offers a reasonably good description of
nucleation when $n >> 1$.  In the case where $n \rightarrow 1$, the
more exact \textit{atomistic} model is required
\begin{equation} \label{eq:nucl_gex}
G_{ex,n} = \lambda n - E_n
\end{equation}
where $n\lambda$ is the work necessary to transfer $n$ particles from
the old phase to the bulk of new one, and $E_n$ is the binding energy
of the particle within the cluster. This form of $G_{ex,n}$ is valid
for all $n$. However the difficulty of the theory is the determination of
$E_n$, which is the result of complex chemical and quantum mechanical
interactions within molecule. The focus of this work in section
\ref{sec:dft_results} was the evaluation of this term. The capillary
approximation discards these difficulties in favor of a simple
droplet model.

Eq.(\ref{eq:nucl_gex}) allows the WCF to be expressed more precisely
in the atomistic form
\begin{equation} \label{eq:nucl_wcf1}
W_n = -n(\Delta \mu - \lambda) - E_n
\end{equation}
To good approximation, $\lambda$ can be treated as an intensive
property of the material. As $\Delta \mu$ is given in terms of the
thermodynamic variables ($T$, $S$) of the system, this leaves the
binding energy $E_n$ as the only unknown quantity in determining the
WCF.  Our DFT calculations allow us to compute $E_n$ directly for
small clusters, without the need to recourse to the capillary
approximation.

The classical nucleation rate (Eq. \ref{eq:nucl_jrate0}) is given in
terms of the surface area and WCF of the cluster at the
\textit{critical size} $n^*$.  The value of $n^*$ is the size at which
the $W_n$ is maximum.  Clearly, determination of the critical size
$n^*$ is essential.  In the capillary approximation $W_n$ is a smooth,
continuous function of $n$, and determination of $n^*$ is accomplished
with elementary analysis. However, in the case of
Eq.(\ref{eq:nucl_wcf1}) the $W_n$ have no known analytic form, and
determination of $n^*$ must be done by direct inspection of available
values.

A more concise form of Eq.(\ref{eq:nucl_jrate0}) can be given using the
monomer attachment rate for a cluster of size $n$ as
\begin{equation}
f_n = \gamma S_n C_1 \sqrt{\frac{kT}{2 \pi m_0}}
\end{equation}
and the $n$-cluster concentration as
\begin{equation} \label{eq:nucl_cons0}
C_n = C_1 e^{-W_n/kT}
\end{equation}
The steady-state nucleation rate is then
\begin{equation} \label{eq:nucl_rate_clconcise}
J_s = z f^* C^*
\end{equation}

The Zeldovich factor $z$ is is present to account for back-decay of
the supercritical clusters.  In the classical theory, it takes the
approximate form
\begin{equation} \label{eq:nucl_cap_z0}
z = \left[\left(\frac{-1}{2 \pi kT}\right)\left(\frac{d^2 W}{dn^2}\bigg|_{n=n^*}\right)\right]^{1/2}
\end{equation}

Evaluation using the capillary approximation from
Eq.(\ref{eq:nucl_gex_cnt}) gives
\begin{equation} \label{eq:nucl_cap_z1}
z = \sqrt{\frac{W^*}{3 \pi kT n^*}}
\end{equation}

In the atomistic case, the derivatives in Eq.(\ref{eq:nucl_cap_z0})
are ill-defined.  Recourse to the kinetic formulation (see
\citealt{LewisTB}) shows that $z$ can be given in terms of $C_n$
\begin{equation}\label{eq:nucl_kin_z0}
z = \left(\sum_{n=1}^h \frac{C^*}{C_n}\right)^{-1}
\end{equation}
where $h > n^*$ is a reasonable cut-off of the summation (see below).
%
%

\subsection{Kinetic nucleation theory}

A more precise treatment of nucleation is given by the kinetic theory,
first formulated by Becker \citep{Becker35}.  Clusters are assumed to
form in a bath of abundant monomers, their size changing by no more
than one atom at a time.  The steady state nucleation rate is given in
terms of attachment and detachment rates ($f_n$ and $g_n$, respectively):
\begin{equation}
\label{eq:nucl_jrate2}
J = f_1 C_1 \left[ 1 + \sum_{n=2}^{h} \frac{g_2 g_3 ... g_n}{f_2 f_3 ... f_n}\right]^{-1}
\end{equation}
In this formulation no recourse to thermodynamics is required, and the
simplifying assumptions of CNT can be discarded provided $f_n$ and
$g_n$ are known. In principle, the attachment and detachment rates
must be known at any size $n$ ($h=\infty$). However, if a critical
cluster $n^*$ is defined for the size at which $f_n=g_n$, the
summation in Eq.(\ref{eq:nucl_jrate2}) can be truncated at $h=2n^*$
(e.g. \citealt{KashchievTB}).

The kinetic nucleation theory is readily amenable to adoption of our
DFT-derived cohesive energies since
\begin{equation}
\label{eq:nucl_detach0}
g_n \propto e^{(W_n - W_{n-1})/kT}\propto e^{[\lambda-(E_n-E_{n-1})]/kT}
\end{equation}
A close look to our results, however, reveals that the ratio $f_n/g_n$
keeps oscillating around unity, preventing us to identify the critical
cluster size in the kinetic sense. However, it is possible to avoid
this difficulty by reintroducing the cluster concentrations $C_n$ defined
in Eq.(\ref{eq:nucl_cons0}), since
\begin{equation}\label{eq:nucl_cons1}
C_n=C_1e^{W_n/kT}=C_1\frac{f_1f_2f_3\dotso f_{n-1}}{g_2g_3g_4\dotso
  g_n}
\end{equation}
Eq.(\ref{eq:nucl_jrate2}) takes on the simple form
\begin{equation} \label{eq:nucl_jrate3}
J_s = \left(\sum_{n=1}^h \frac{1}{f_n C_n} \right)^{-1}
\end{equation}
Inspection of the above reveals a clear dominance of the term at
$n=n^*$, where $n^\star$ is here defined as the value of $n$ that
maximized $W_n$, as in CNT.  Truncating the summation at $h=2n^*$ has
negligible effect on the result.

Values for bulk sublimation energy $\lambda$, surface tention $\sigma$, and carbon molecule mass $m_0$ are those used in \cite{Fallest12}.
\subsection{Results}
\subsubsection{Work of cluster formation}

\begin{figure}
\epsscale{1.0}
\plotone{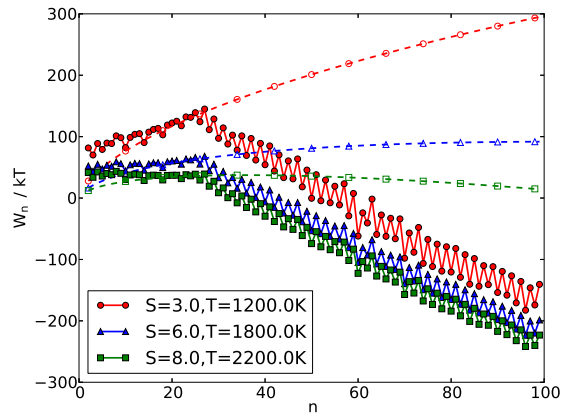}
\caption{Work of
    Cluster Formation for three values for temperature, saturation.
    Solid curves are from this work, dashed curves are $W_n$ in CNT.
    $W_n$ given per $kT$}
\label{fig:wcf_0}
\end{figure}

Fig.(\ref{fig:wcf_0}) shows the WCF for different values of
temperature and saturation in both the CNT and the accurate atomistic
cluster formation energy using the results of DFT calculations.  The
overall trend of the atomistic evaluation changes nearly linearly on
both sides of the fullerene transition. Higher temperatures and
saturations give a much slower growth before the fullerenes. It is in
this regime where CNT is most near the atomistic case. The classical
WCF coincides most closely at the high-$n$ ring structures.  CNT
estimates lower WCFs at the low chains.

The most striking divergence between CNT and the atomistic case is at
the fullerene transition.  Here the is a distinct change in the trend
of the cluster energies.  At low temperatures and saturations this is
clearly the critical size.  At higher temperatures and saturations, the
critical size is located at a lower cluster size, although a distinct
transition into the fullerenes is still apparent.  This change arises
from the properties of the fullerene in relation to the previous
geometries. In particular, the fullerene is 3-coordinated (compared to
the 2-coordinated rings) and addition carbon atoms releases more
energy than in the 2-coordinated case.  As seen below in the results
for the critical cluster size, the fullerene transition (and, to a
lesser extent, the ring transition) provide a natural critical size
past which growth in the new phase is spontaneous.

\subsubsection{Critical cluster size}
\begin{figure}
\epsscale{1.0}
\plotone{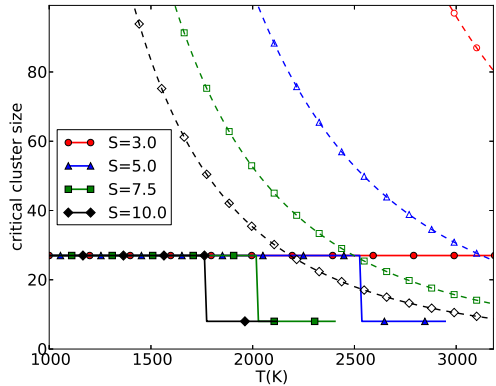}
\caption{Critical cluster
    size $n^*$ as a function of temperature at different values of
    saturation. Solid curves are results of this
    work, while dashed curves and symbols are derived from CNT.}
\label{fig:ncrit_0}
\end{figure}

Critical cluster sizes in the classical and precise atomistic case are
given in Fig.(\ref{fig:ncrit_0}). Critical sizes in both cases are
determined by selecting the value of $n$ for the maximum value of the
WCF, if such a maximum exists for $2 < n < 99$.  CNT gives
every-increasing values for the critical size as the temperature
falls. In the atomistic case the maximum critical size at $n^*$ is
consistent at $n^*=27$ for a wide range of temperature and
saturation. The critical size falls in both models with increasing
temperature, however the atomistic case gives a minimum critical size
of $n^*=8$.

At higher temperatures and saturations, the WCF is maximum at $n=2$,
implying that a catastrophic seeding of the new phase should take
place. In practice, since at high temperatures nucleation is fast, the
gas phase is depleted at very small saturation when $n^*>2$, and
saturation never reaches values for which catastrophic nucleation
occurs.

\subsubsection{Nucleation rates}

As discussed above, we will adopt three methodologies to compute the
nucleation rate. In the first method, named CNT-CAP we adopt the
capillary approximation within the classical nucleation
theory. Nucleation is computed from Eq.(\ref{eq:nucl_jrate0}) with the
Zeldovich factor from Eq.(\ref{eq:nucl_cap_z1}), surfaces calculated for
spherical uniformly filled grains, and WCF from the capillary
approximation Eq.(\ref{eq:nucl_gex_cnt}). The second methodology,
named ANT-DFT still adopts the classical nucleation theory, but the
capillary approximation is replaced by our DFT results on the
clusters' cohesive energies. The WCF is computed from
Eq.(\ref{eq:nucl_wcf1}), and the cluster surfaces derived from the most
stable configurations in DFT.  In practice, we compute the surfaces
with the formula
\begin{align}
S_n &= 2 \pi r_C n l + 2 \pi r_C^2\\
S_n &= 2 \pi r_C (n - 1)l\\
S_n &= 12 A_p + (n/2 - 10)A_h
\end{align}
for the chain, ring, and fullerene \citep{Adams94} areas respectively.
$l$ is an assumed bond length, and $r_C$ is the carbon atom radius.
We take as $l = 1.4 \angstrom$ and $r_C = 1 \angstrom$.
The Zeldovich factor is computed atomistically using Eq.(\ref{eq:nucl_kin_z0}).

Finally, we compute the nucleation rate with the full kinetic theory,
and we label these results as KNT-DFT. The nucleation rate in this
case is computed with Eq.(\ref{eq:nucl_jrate3}), where the cluster
concentrations are derived according to Eq.(\ref{eq:nucl_cons0}) and
making use of our DFT-derived cohesive energies.

\begin{figure}
\epsscale{1.0}
\plotone{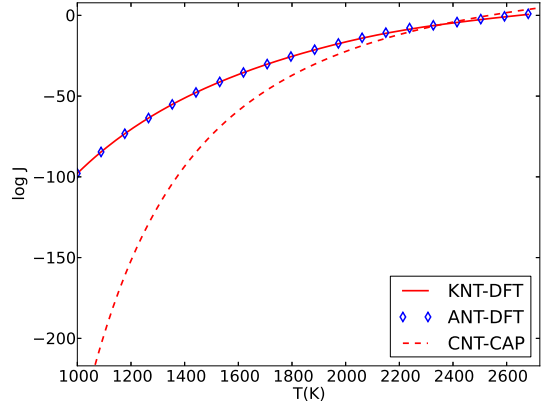}
\caption{Nucleation rates for a single saturation using CNT-CAP (dashed), ANT-DFT (diamond), \& KNT-DFT (solid).}
\label{fig:jrate_1}
\end{figure}

\begin{figure}
\epsscale{1.0}
\plotone{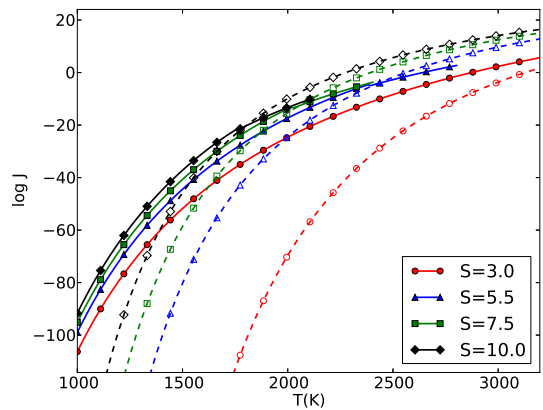}
\caption{Nucleation rates at different saturations using ANT-DFT(solid, filled) and CNT-CAP (dashed, open).}
\label{fig:jrate_0}
\end{figure}

A comparison of CNT-CAP, ANT-DFT, and KNT-DFT nucleation rates is
plotted in Fig. ~\ref{fig:jrate_1}. ANT-DFT and KNT-DFT rates agree to
very high order.  Expanding Eq.(\ref{eq:nucl_rate_clconcise}) with
Eq.(\ref{eq:nucl_kin_z0})
\begin{align*}
J_{cl} = z f^* C^* &= \left[\sum_{n=1}^h \left(\frac{C^*}{C_n}\right)\right]^{-1}f^* C^*\\
&=\left[\sum_{n=1}^h \left(\frac{1}{C_n}\right)\right]^{-1} f^*\\
&=\left[\sum_{n=1}^h \left(\frac{1}{f^* C_n}\right)\right]^{-1}
\end{align*}
Cleary this only differs from Eq.(\ref{eq:nucl_jrate3}) by the the
constant $f^*$ in the summation.  The $f_n$'s vary much more
slowly than the $C_n$'s, and the difference between the two sums is
negligible.

Since the ANT-DFT and KNT-DFT results are in good agreement with each
other, we only show in the following the results of ANT-DFT. This is
not a trivial result and it is particularly useful for numerical
implementations of nucleation, since the prescription of ANT-DFT is
much faster than KNT-DFT, especially at large critical sizes.

ANT-DFT rates and CNT-CAP rates are compared for a set of saturations
in Fig.(\ref{fig:jrate_0}). Nucleation at lower temperatures is faster
than in the capillary model, due to the much smaller critical size
given by DFT. As the critical size predicted by the two calculations
converge the rates of the methods track better.  At large temperatures
and saturations nucleation is suppressed.  In this regime DFT predicts
no critical size beyond the dimer. Transition between the two phases
is spontaneous.  Lower saturations allow higher nucleation
temperatures.

\section{Conclusions and Discussion}
\label{sec:conclusion}
We performed empirical global geometry optimization on carbon clusters
$n=2$ to $n=99$ using a combination of search algorithms to find
lowest-energy candidate clusters. We then applied local geometry
optimization and binding energy calculation using the DFT code {\it
  Quantum Espresso} to determine the precise ground state.  These
results are applied to accurate atomistic nucleation theory and
compared with the results of classical nucleation theory.

Where there is overlap, our results for the geometries and energies
conform well to previous studies. We find good
agreement with calculations performed using similar DFT methods (see
Table \ref{tbl:compare}). Higher energies than quantum Monte Carlo
techniques are found, although we find a similar ring-to-fullerene
transition pattern \citep{Kent99}. As expected, our results for small
clusters are consistently lower than corresponding LDA methods, which
are known to overestimate energies and bond strengths. All DFT methods
overestimate the energies compared to experimental values for small
clusters.

Low-$n$ clusters are mostly chains, with C$_6$ the only ring structure
due to the strong stability of Huckle aromaticity. Carbon rings begin
to form at C$_{10}$ and continue up to C$_{27}$. These rings are
initially distorted and show lower symmetry than their counterparts in
classical evaluations, but this distortion declines with larger $n$.
Fullerene and fullerene-like structures start at C$_{24}$, and form all ground state
configurations from C$_{28}$ to C$_{99}$.

Even though the calculations were performed at zero pressure, they are
expected to provide an excellent approximation for the typical
pressures at which astrophysical nucleation takes place (less than one
barye). As an example, consider the $n=92$ fullerene. It can be
re-casted in an onion structure with an $n=32$ fullerene inside a
$n=60$ buckminsterfullerene. This nested structure has a lower total
binding energy by approximately $10$~eV or
$1.9\times10^-{11}$~erg. Since the volume of the C$_{92}$ fullerene is
of order $10^{-22}$~cm$^2$, a very large pressure of approximately
$10^{11}$~barye is required so that the more compact onion structure
has a lower enthalpy then the bigger C$_{92}$ fullerene.

Critical cluster sizes and nucleation rates were calculated with these
results and compared to the rates obtained with the classical
capillary approximation. We find that the critical cluster size
consistently falls near the two geometry transitions (chain-to-ring,
ring-to-fullerene), while it varies continually as a function of
temperature and saturation adopting the capillary approximation. As a
consequence, our ANT-DFT results yield larger nucleation rates at low
temperature and saturations, the regimes in which CNT-CAP predicts
very large critical clusters. In support of our conclusion, previous work
has shown that ANT offers a richer and more natural nucleation description 
than CNT by accounting for irregular cluster geometries \citep{Kashchiev12}.

Nucleation rates at higher temperatures
and saturations tend to be similar, as the difference in critical
cluster size between the exact case and the capillary approximation
becomes smaller. An important result particularly relevant for
numerical implementations is that the nucleation rates do not depend
on the use of the classical or kinetics frameworks. The more
computational intensive kinetic theory yields rates only marginally
different from the classical thermodynamic framework, as long as the
work for cluster formation, the cluster surfaces, and the Zeldovich
factor are computed with DFT cohesive energies instead of with the
capillary approximation.

\acknowledgements This work was supported in part by NSF grants
1150365-AST and 1461362-AST (DL \& CM).  We thank the Texas Advanced Computing Center
(TACC) at the University of Texas Austin for providing HPC resources.

\clearpage

\end{document}